# Multimodality in Group Communication Research

Robin Lange*, Brooke Foucault Welles*, Gyanendra Sharma*, Richard J. Radke**, Javier O. Garcia***, & Christoph Riedl*

*Northeastern University

** Rensselaer Polytechnic Institute

*** US DEVCOM Army Research Laboratory, Humans in Complex Systems Division

## Introduction

Generations of research assistants have toiled for countless hours to manually code eye movements, hand gestures, speech, and other minutia of human communication to facilitate the study of groups and teams. Manual coding is a time intensive, mentally draining, and sometimes subjective and error prone procedure in which a researcher must code the slightest switch in attention, spoken syllable or miniscule movement. Recently, technology has begun to provide coding assistance in the form of machine learning and AI. With technological assistance, transcription of verbal input can now be done in a quarter of the time that it would take human coders (Bazillon et al., 2008). The speed, or lack thereof, of the manual transcription process has historically meant that studies that collect data through sensors, such as video or audio capture technology, are time intensive and expensive to produce and this cost often doubles or triples for group research. Furthermore, researchers who wanted to study multisensory processes, such as the interplay between eye gaze and speech, were largely unable to do so due to the prohibitive cost of multiple sensors to capture data, and human coding to transcribe and annotate it for analysis. This means that for much of the history of group communication research, most studies were constrained to a single modality, using data from only one sensing technology such as just audio, or was incredibly expensive to produce. With the advances in modern technology, including multimodal data capture and machine learning processing algorithms, we now have the ability to conduct multimodal research, research that combines multiple modalities, to capture the multisensory process of human interaction. This promises to unlock novel insights and opportunities for theory development that will lead to a better understanding of interactive and higher order group processes such as leadership that emerge from an interplay of verbal, para-verbal, and non-verbal group communication. For example, we know that emergent leaders tend to be the center of visual attention while speaking (Sanchez-Cortes et al., 2013) and also make inviting gestures through eye gaze at the end of a spoken statement to invite others to speak (Kalma, & Van Rooij, 1982).

### Why should Group Research be Multimodal?

In group meetings, people continuously act and react to each other to communicate and achieve shared goals. Dynamic group communication works because most people are adept at reading and responding to the complex tapestry of verbal, para-verbal, and non-verbal communication such as body language, facial expressions, and other behavioral markers that comprise social interaction in groups (Kern & Tindale, 2014). By extension, studies of communication in groups are incomplete when they only study one sensing modality. *Multimodal data*, or data collected using two or more sensing modalities, are crucial to study the multiple (often simultaneous) behaviors typically expressed when people communicate, especially when they communicate in groups. Multimodal research conducted on groups is incredibly useful for identifying higher order properties such as leadership. For example, research using multimodal data has been able to identify emergent leaders (Sanchez-Cortes et al., 2013) better than unimodal research (Bhattacharya et al., 2018) and has been able to identify leadership style (Beyan et al., 2017b). A multimodal research approach promises a more complete understanding of group communication processes and outcomes. This promises to significantly advance communication and management theory. Finally, reducing costs and automating data collection opens opportunities to study more (and possibly larger) groups which can lead to richer theory by enabling researchers to study heterogeneity across individuals and contexts (cf., Anderson, 1972).

## Technology and Multimodal Research

Multimodal research relies heavily on modern sensing technology and machine augmented coding methods. Before modern technology, human coders needed to extract interaction processes such as speech or gesture in situ. With advancing technologies, we now have sensors that are able to capture this information for us. We can use microphones to capture audio, cameras to capture visual information, depth sensors to capture position (and ultimately gestures), and even biosensors to capture neurophysiological information about participants. We can now (relatively) easily look at interactive multimodal processes like eye gaze and speech, gesture and heart rate, or EEG and body language. We can then use signal processing and machine learning algorithms to detect patterns previously only recorded by human coders (Brauner et al., 2018; Luciano et al., 2018; Pilny et al., 2019) in fractions of the time. Technology can augment, or in some cases entirely replace, human coding which has significantly expanded the possibilities of group research. This opportunity means that research that was formerly impossible is now possible. Technology-enabled multimodal research has readily been embraced by researchers in neuroscience, biological engineering, human-computer interaction, and related fields. Yet, it is unclear whether communication and management scientists are using these same opportunities.

This paper examines the state of multimodal research in communication and management research. We proceed in three parts. First, we provide a definition of what we mean by multimodal communication research and introduce different modalities and their capabilities. Second, we present a systematic literature review to demonstrate a scarcity of multimodal group communication studies and highlight opportunities for combining single-mode studies. Third, we offer instrumentation suggestions that explain how technology can be used for data collection. Finally, we offer suggestions of analysis and data integration methods to support a multimodal research process including ethical considerations.

## Part 1: What we Study in Multimodal Studies

For the purposes of this paper, we focus exclusively on synchronous communication in groups. This includes group communication that is either face-to-face or computer-mediated video and/or audio. In this section we note the different behavioral groups of interest and their features.

**Verbal Behavior**

Verbal behavior is one of the most informative and important cues during group interactions (Danescu-Niculescu-Mizil et al., 2012). For example, in meetings with performance objectives, verbal communication is integral to reaching such goals (Littlepage et al., 1997). Verbal behavior maps onto the audio modality and is captured by audio recording devices such as microphones. See Figure 1 for an overview of the behavioral metrics in this modality as well as methods of capturing this data.

The verbal behavior of participants in groups can be analyzed to extract various key behavioral markers for individuals and coordinated markers of groupwork. For example, there is well-established evidence that leaders in groups tend to display key verbal behaviors such as topic initiation, course correction, and concluding remarks in abundance (Gerpott et al., 2019; Klonek et al., 2018; Sudweeks & Simoff, 2005). Other interactive behaviors such as dominance and performance in meetings have also been shown to have significant relationships with key behavioral markers that are derived from the verbal class, such as competency, sentences spoken, affective language, language synchronization, mimicry, and interruptions (Anderson & Kilduff, 2009; Huffaker, 2010; Gonzales et al., 2010; Woolley et al., 2022).

Verbal communication is not limited to speech but also includes para-verbal/prosodic behaviors. Such behaviors include tone, pitch, interruptions, and pacing of verbal communication. Paraverbal behaviors are linked to various behavioral features such as participation, emphasis, curiosity, and engagement, and even personality traits such as dominance (Dubey et al., 2017; Pianesi et al., 2008). These behaviors are widely accepted as reliable social markers when studying group dynamics and behavior (Remland, 1984). Communication within groups is largely conditioned by non-verbal parameters of speech, where prosodic expressions signal perceived hierarchical orderings among the group members. For instance, dominant individuals tend to show higher levels of energy and variance in intonations (Mathews & Braasch, 2018). Beyond hierarchical orderings within a group, acoustic and prosodic features have shown high degree of correlation to social behavior (Gravano et al., 2011). Similar verbal content expressed with differences in intonation, pitch, or energy can imply entirely different connotations related to behavior and personality traits (Mohammadi & Vinciarelli, 2012). Thus, standalone analysis of the verbal content of a group meeting without considering prosodic features may lead to misleading or erroneous conclusions.

**Non-Verbal Behaviors**

Communication in synchronous groups (in-person or remote) is shaped through a combination of verbal and non-verbal features such as body language, head, and body orientation, arm movements, and eye contact. Non-verbal communication provides important real-time feedback to the speaker. For example, making eye contact with an active speaker is largely considered a sign of attentiveness or interest towards the content of the speech (Lucas et al., 2016; Zhang et al., 2020). Non-verbal behaviors can also give researchers insight into processes like leadership. In leader-follower situations, leaders garner significantly more attention, expressed via behaviors such as eye contact, visual focus of attention, shared gaze, etc. (Gerpott, Lehmann-Willenbrock, Silvis, et al., 2018; Gerpott, Lehmann-Willenbrock, Voelpel, et al., 2019; Sanchez-Cortes et al., 2013). Non-verbal behaviors are often captured by 2-D or 3-D sensors (see Figure 1).

**Covert Physiological Processes**

The complex processes underlying communication in groups requires a combination of both neural and non-neural physiological processes (Ochsner, 2004). We use the term *covert physiological processes* to encompass all physiological dynamics, including but not limited to the gamut of neural signals from a variety of technologies that span spatial and temporal scales (Bassett & Sporns, 2017), and non-neural signals including cardiovascular and dermal measurements (Heikenfeld et al., 2018; Imani et al., 2016), see figure 1. These covert physiological processes can give us insight into phenomena underlying aspects of communication, such as information sharing (Baek, et al., 2021; Lima Dias Pinto et al., 2022; Doré et al., 2019), risky behavior (Kim-Spoon et al., 2017), and the tendency to mirror behavior (Wasylyshyn et al., 2018). Understanding covert physiological processes has the potential ability to drastically improve our understanding of organizational behavior.

**Part 2: Multimodality in the Existing Literature**

Despite the promise of multimodal research for understanding group communication, it is unclear how much research utilizes multimodal approaches. We carried out a meta-study of published studies in group communication and management to survey how many studies were multimodal, second to identify common modalities and multimodal combinations, and finally to examine algorithmically augmented analysis techniques.

**Data**

To establish a corpus of research articles on group communication we searched the ISI Web of Knowledge Social Sciences Citation Index (SSCI) database using the keywords "group" or "team" and "behavior coding" (TS = (group* OR team* AND behavior coding). The search was limited to journal articles published between 2005 - 2020). Review papers were excluded. This resulted in a primary database of 24,877 scholarly articles.

We further filtered based on the publication venue and only consider those that were published in premier management and/or communication journals. This process yielded a set of 2,316 scholarly articles. From this set of papers, we selected all articles with at least 10 citations, resulting in a total of 1,474 articles with some scholarly impact. Selecting for papers with citations negatively affects recent publications. So, we also included all of the papers published in the last two years –2019 - 2020). This added an additional 221 papers that were not already included based on citations. In total, 1,691 papers were included in our meta study.

**Coding**

All papers were coded on three categories: whether they were empirical research, whether they were multimodal, and if they used algorithmic or computational methods (signal processing or machine learning). Only studies that reported empirical research were included for subsequent analysis. The remaining papers (literature reviews, commentaries, or computer modelling papers) were excluded (439 papers).

For this study, we grouped the behavioral features into modalities and then mapped those modalities to different sensors. For example, we grouped verbal and paraverbal behavior to the audio modality and then mapped the audio modality to audio sensors such as microphones. Papers were then coded for their inclusion of audio, visual, and/or biosensing sensors if they explicitly stated they used the use of that capture method. Static capture encompassed any data that was captured at one time point, such as surveys. Audio capture encompasses all data captured using sound, visual capture encompasses all data using light and depth, and biosensing capture encompasses all neurophysiological data. In practice, this meant that the use of microphones counted for audio capture, video for 2D capture, depth sensors for 3D capture, and sensors that record physiological data, such as heart rate monitors, were coded as biosensing capture. Video cameras such as webcams also typically record sound, so their use was also counted toward audio capture, unless the paper explicitly stated otherwise. Studies were considered multimodal if they used two or more of these capture technologies. Papers were coded as using machine learning methods if this was explicitly mentioned in the paper. In situations where a paper combined multiple studies, each study was coded and analyzed as a separate entry. See Appendix 2 for the full coding guide.

Data was coded by four independent raters. Interrater reliability was analyzed on the study type category. The Krippendorff's alpha of all raters was above 0.86 pooled across all coders. Between the pairs of coders, Krippendorff's alpha was between 0.75 and 1.

## Results

| Type of Research | Number of Studies | Number of Multimodal Studies | Number of Papers | Number of Unimodal Papers | Number of Multimodal Papers |
|---|---|---|---|---|---|
| Empirical Research | 1590 | 208 (13% of the Empirical Studies) | 1255 | 1075 | 201 |
| Commentary | 323 | | | | |
| Review Paper | 74 | | | | |
| Meta-Analysis | 31 | | | | |
| Computer Modeling | 12 | | | | |

Table 1: Papers and Studies by Type of Research and Modality

Most papers in our sample were empirical research (1,255 papers). These papers contained 1,590 studies. On average, each paper contained 1.3 studies. Unsurprisingly, there were few review papers and meta-analyses in our sample. Some papers contained both unimodal and multimodal research. These papers counted both as unimodal and multimodal papers. There were 1,426 unimodal papers and 208 multimodal papers.

| | Number of Multimodal Studies | Number of Multimodal Papers |
|---|---|---|
| Audio Sensors | 200 | 194 |
| 2-D Sensors | 60 | 57 |

| 3-D Sensors | 7 | 3 |
| --- | --- | --- |
| Biosensing Sensors | 0 | 0 |
| Machine Coding | 15 | 12 |

Table 2: Modalities broken down into Studies and Papers

In total, 1,590 studies were coded; of these studies, 201 (12.6%) were multimodal and 1,389 (87.4%) were unimodal. In our analysis, the use of capture technology was used as a proxy for each of the modalities. Other than the static modality, most of the multimodal studies used at least audio capture; fewer used 2-D and 3-D sensors and none used biosensing sensors.

Some combinations of research were more common than others. Because the static modality was the most commonly combined with the other modalities, it has been excluded from the following counts. Of the non-static modalities, audio and 2-D capture technology were the most common to be combined (56 studies), followed by audio and 3-D capture studies (4) and 2-D and 3-D studies (1). There were no studies that used the audio, 2-D and 3-D sensors.

Audio sensors can capture both verbal and paraverbal behaviors such as topic initiation, course correction, concluding remarks in abundance and energy and variance in intonations and interruptions (Gerpott et al., 2019; Klonek et al., 2018; Mathews & Braasch, 2018; Sudweeks & Simoff, 2005; Woolley et al., 2022). 2-D sensors can record many nonverbal behaviors such as body language, eye-gaze and body orientation. Some 2-D sensors can do poorly at capturing depth information, which is where 3-D sensors excel. 3-D sensors can capture depth information and when all participants are wearing sensors, the relative position of many participants simultaneously. Biosensing sensors can capture covert physiological processes such as heart rate or neural signals.

The multimodal studies were further examined to see which studied the multisensory interactions of the different modes. While some studies used multiple sensor types, none looked at the interactions and complexities that exist at the combination of multiple modes of study. This means that though a study may have used video cameras and audio recording devices they did not look at the possible interactions within the data from both sources.

There were 18 studies (15 papers) in our sample that used machine learning methods. Of these studies, all were multimodal and none were unimodal. This means that only 8.7% (18/208) of the multimodal studies in our sample used machine learning methods.  Of the studies that use machine learning methods, 10 examined audio only, 1 examined an audio and 2-D together, 4 examined audio and 3-D, and 1 examined 3-D.

We used linear regression to test if papers that use multimodal research methods garner the same number of citations as unimodal research. We find no significant difference between single vs. multimodal methods and citations of an article (controlling for journal and year fixed effects; $F(38,1593) = 123.7$, $p = 0.096$; $R^2 = 0.74$).

**Publication Venues**

While still relatively uncommon, many venues are receptive to multimodal work. In total, 22 journals in our sample published empirical research. See Appendix 1 for a full list of journals in our sample. Of these journals, only one did not publish any multimodal research. The journal that published the most multimodal research from our sample was the Academy of Management Journal (26 papers), followed by the Journal of Applied Communication Research (25), Organization Science (25), the Journal of Applied Psychology (17), and Organizational Behavior and Human Decision Processes (17). In comparison, the journals in our study that published the most empirical research were the Journal of Applied Psychology (207 papers), the Academy of Management Journal (134), Organizational Behavior and Human Decision Processes (117) and Organization Science (114). Proportionally, the journals that published the most multimodal research in comparison to the other papers in our sample were Narrative Inquiry (78%; 7/9 papers), Journal of Applied Communication Research (61%; 25/41), Visual Communication (33%; 1/3), and Administrative Science Quarterly (32%; 15/46).

## Discussion

From the meta-study it is clear that multimodal research is utilized very little in the communication and management literatures. Even when studies use multiple sensors, they are not looking at the dynamic interplay of the senses that humans are picking up on implicitly. Improvements in sensor technology and analysis techniques (e.g., machine learning) are enabling multimodal group research for a broader range of researchers. However, from the meta-study, we have learned that in top communication and management journals, most research on group interaction is unimodal and does not take advantage of potential multimodal options. Of the papers that use multiple capture technologies, most use the audio modality and the 2D vision modality. This is unsurprising because it is likely that many researchers already have experience with these modalities and thus there is a relatively low barrier of entry. In addition, there are many useful phenomena that can be captured by the audio and 2D vision modalities.

To our surprise, even when studies used multiple sensors, they were not looking at the interactions that happened across the modalities captured by those sensors. There could be a few possible reasons for this. First, though an implicit part of much of human interaction, researchers might not be thinking of the ways that interaction is multisensory. They might implicitly understand speech and gesture are linked but be unable able to explicitly recognize the importance of multisensory interactions. However, as previously stated, understanding these interactions gives us a better understanding into processes like leadership, because leaders use gestures through eye gaze at the end of a statement for other participants to speak (Kalma, & Van Rooij, 1982). Another potential reason is that researchers are interested in these processes but are unprepared to integrate the different data streams into a singular analysis. Data from multiple streams need more consideration and preparation, for example, in the use of a synchronizing step (for more detail see Part 4).

No studies in our sample used the biosensing modality. These modalities appear to be the purview of studies focused on the underlying mechanisms of communication, team processes, and social neuroscience, and are published in other outlets such as Neuroscience and Biomedical Engineering Conferences and Journals (for example, NeuroImage, Network Neuroscience, Journal of Neuroscience, etc.). There have been several advances in wearable technologies to sense different physiological processes and in the unobtrusive passive monitoring of neural processes. These technologies are often lightweight, easy to instrument on individuals, and there exist several off-the-shelf options for accurate synchronization such as lab streaming layer, a software solution that allows for the synchronization of physiological and non-physiological streams of information (Kothe, 2014). Given these advances in biosensing technology that make it feasible for group research, it is unclear whether the barrier to entry is difficulty with technology, difficulty finding skilled collaborators, or a lack of interest in using such approaches. The former two seem most likely. Waldman (2013) argued that regarding neuroscience in group research, organizational researchers who likely lack experience in neuroscience and are interested in this field of study would benefit from neuroscience collaborators. This hurdle, while not insurmountable, may contribute strongly to the lack of biosensing papers in our sample.

To help advance the use of technology to enable multimodal research in Communication, Management, and related social sciences, in the remainder of this paper, we outline equipment and other considerations for researchers planning multimodal group research studies. We offer instrumentation suggestions mapped to specific modes of interest, multisensorial data collection and analysis guidelines, and reflections on the limitations and risks of technology-augmented multimodal group research, paying particular attention to the robustness of analysis methods, including grappling with known algorithmic biases.

## Part 3: Multimodal Signals and Instrumentation for Data Collection

Multimodal research has been enabled, by advancing sensing technologies and machine learning analysis techniques. The advent of sensing technology resulting in rich, multimodal data streams, along with emerging machine-learning-based analysis algorithms, allows researchers to study increasingly complex dynamic behaviors. The field is inherently interdisciplinary, requiring expertise in hardware (sensors), software (machine learning), data science, and social science. The challenge of combining insights from different fields to advance group research can explain why the area is still in its infancy. Here, we group communication behaviors into data types and from data types into sensors (see Figure 1). This section covers an overview of some technological considerations researchers should take into account when planning a multimodal research study.

| Behavioral Mode | Sensors | Example Devices | Example Group Communication Processes |
|---|---|---|---|
| Verbal Behavior | Audio | • Lapel Microphones (Micro)<br>• Microphone Array (Meso)<br>• Common Microphone (Meso) | • Topic Initiation<br>• Course Correction<br>• Concluding Remarks<br>• Competency (TF-IDF, informativeness)<br>• Full Sentences Spoken<br>• Synchrony<br>• Speaking time |
| Paraverbal Behavior | | | • Affect<br>• Interruptions<br>• Tone<br>• Pitch<br>• Vocal Relaxation<br>• Backchannels<br>• Sound Signal Energy |
| Non-Verbal Behavior | 2-D Sensors | • Individual Cameras (Micro)<br>• Webcam Camera Array (Micro)<br>• Single Video Recorder (Meso) | • Eye Gaze/ Eye Movement<br>• Mouth Movement<br>• Shared Gaze<br>• Attention Direction (Giver, Receiver, Center)<br>• Body language |
| | 3-D Sensors | • Overhead Depth Sensors (Meso, Micro)<br>• Face Capturing 3-D Sensors (Micro) | • Location<br>• Orientation<br>• Seating Posture (Forward, Backward, etc)<br>• Arm Posture<br>• Gesture |
| Covert Physiological Processes | Neuro-Physiological Sensors | • Heart Rate Monitors (Smartwatch, Fitbit, etc)<br>• fMRI<br>• EEG<br>• EMG<br>• Epidermal Sensors | • Heart Rate<br>• Cognitive Performance<br>• Stimuli Neurological Responses<br>• Epidermal Sweat |

Figure 1: Sensor modalities within the scope of multimodal analysis for group research. There are four different sensor types, each of which correspond to either one or multiple behavioral modes that can be studied as part of group interactions. Please note that the examples of communication processes are nonexclusive to each sensor type but are grouped by ease of capture by sensor. For example, location data is visible with 2-D sensors but is grouped under 3-D sensors because 3-D sensors can excel at capturing location data.

**Capturing Verbal/Audio Data**

Verbal behavior is related to the content of speech and requires a two-step process to capture: recording and transcription. Recording is a relatively straightforward process in which experimenters record communication, though design considerations can significantly vary depending on whether the objective is to study group dynamics at the individual (micro) or group (meso) level. Studies that require micro analysis should consider deploying individual recording devices per participant, whereas for a meso level analysis, one recording device for the entire meeting would suffice (Figure 1). Table 3 covers an overview of some of the current options that are available.

| Type of Microphone | Benefits | Drawbacks |
| --- | --- | --- |
| Single Microphone for the entire group | Cheapest Option | Difficult to separate multiple speakers (speaker segmentation), by human coders and signal processing algorithms |
| Individual Lapel Microphones for each participant | Cleaner recordings<br><br>Successfully deployed for groups of eight people (McCowan et al., 2005). | More expensive<br><br>Increase salience of being recorded |
| Spherical Microphone Array | Can record up to 16 channels<br><br>When used with beamforming techniques applied to focus on speech from each individual speaker (Mathews & Braasch, 2018) this allows for speaker segmentation in real time | More expensive<br><br>This technology is constantly evolving |

Table 3: Audio Recording Devices Overview

**Transcription**

Once verbal and audio data are captured, it needs to be transcribed for analysis. In the early groups research, transcription was largely a manual process. Today, automatic transcription services, both open source and enterprise (e.g., Google Speech to text, IBM Speech to text, CMU transcription toolkit, Stanford NLP tool), have improved a great deal in the last two decades and offer highly accurate results.

In most cases, the output of such services still requires some manual correction for errors which may require significant human effort. The extent of acceptable inaccuracies largely depends on the research objectives and available resources. It should also be noted that the type of transcription performed by automated transcription services is generally different from that performed by conversational analysts (e.g., Clayman & Gill, 2012; Sidnell, 2012). Typically, automated transcription focuses only on content, not on non-verbal interaction like emphasis, overlapping speech, pitch, intonation, silence, and inhalations.

**Paraverbal Behavior**

The design consideration in regard to sensor-fitting and instrumentation follow the same pathways as for verbal behavior. Paraverbal behaviors include interruptions, back channels, speaking time, turn taking, and social networks based on conversation. Sensors and signal processing algorithms can successfully capture such paraverbal processes at the required frequency (Luciano et al., 2018). Open-source software such as Google's WebRTC Voice Activity Detector (VAD) and packages provided by researchers make the study of paraverbal communication accessible.

## Visual Sensing

In this section, we discuss visual sensing to capture non-verbal behaviors, including both traditional video cameras (2D vision) and newer depth sensors (3D vision).

### 2-D Vision

Video cameras are ubiquitous, inexpensive, easy to deploy, and reliable. Fine-detail interactions, e.g., eye and mouth movements, can be captured using small cameras that are pointed at individual participants' faces. When group size is small (smaller than five), we advise one camera per participant. With larger groups, a wide-angle high-resolution camera can be used to capture multiple participants (e.g., several people on one side of a table). The biggest challenges in studying large groups with multiple cameras are calibration and synchronization, which may require custom software/hardware and technical expertise and preparation. All multimodal researchers (regardless of which modalities they are using) should consider adding in an extra camera in case other sensors fail or misbehave. Since many video cameras also record sound, they are appealing for multimodal research (although the sound quality is generally poorer than dedicated microphones).

### 3-D Vision

Non-verbal means of communication, for instance body language, is expressed via a variety of signals including pose, head orientation, body orientation, hand gesture, arm pose or the distance between participants (Bhattacharya et al. 2018; Chung, 2007). All these signals are significantly easier to determine and interpret at a machine level if depth data is available (Andersen et al., 2012). This information is measured using either time-of-flight (e.g., Microsoft Azure Kinect) or stereo imaging (e.g,. Intel RealSense) technology. Commercial 3-D vision sensors have only come to prominence in the last decade or so, but the resulting richness of data has garnered significant scientific work that can be incorporated into group dynamics research (Bhattacharya et al., 2018; Bhattacharya et al., 2019).

## Wearable and Neurophysiological Sensors

Our discussion so far has mostly been limited to sensors that capture data that can be observed by trained experimenters; however, the scope of group dynamics research could be greatly improved by capturing *physiological processes* underlying group communication. Multimodal imaging combined with other physiological sensors, have allowed an understanding of how neural and non-neural physiological complex systems interact to create complex behavior. This has led to a basic understanding of the placebo effect (Stefano et al., 2001), mindset impacts on stress (Crum et al., 2013), and unexpected biases in perceptions and interactions with the complex world (Azevedo et al., 2017). The scope of possible sensor types spans heart rate monitors to dermal sensors to EEGs (see Figure 1). To detail the considerations for each type of sensor goes beyond the scope of this paper, as each sensor requires different considerations, capture processes and expertise. For this reason, we are also leaving out information on feature extraction and data integration (see Part 4) of neurophysiological signals. Researchers interested in using neurophysiological sensors should consider finding collaborators experienced with neurophysiological sensors and carefully consider both feature extraction and data integration for their multimodal research studies.

**Part 4:** Multimodal Analysis and Data Integration

A generalized multimodal approach to studying group dynamics consists of a multi-step framework that requires some or all of the following aspects: application of various formal algorithms, sound theoretical foundation, questionnaires, qualitative subjective behavioral responses, and objective indicators of performance, communication, or emotion. As shown in the overall data integration/workflow chart in Figure 2, both quantitative and qualitative aspects of group studies are essential to draw reliable conclusions. Machine learning algorithms can automate extraction of objective behavioral features, while manual inputs such as post-task behavioral questionnaires help contextualize and consolidate objective information into broader social and behavioral theories of group dynamics.

In earlier sections of this paper, we focused on sensor instrumentation for multimodal research. Here, we outline the overall framework, incorporating the sensor fusion, quantitative, and qualitative aspects of the research study. We would like to note that while this section is concerned with integration acorss different recording devices, it is also possible to integrate within one class of device, for example verbal and prosodic behavior from audio recordings.

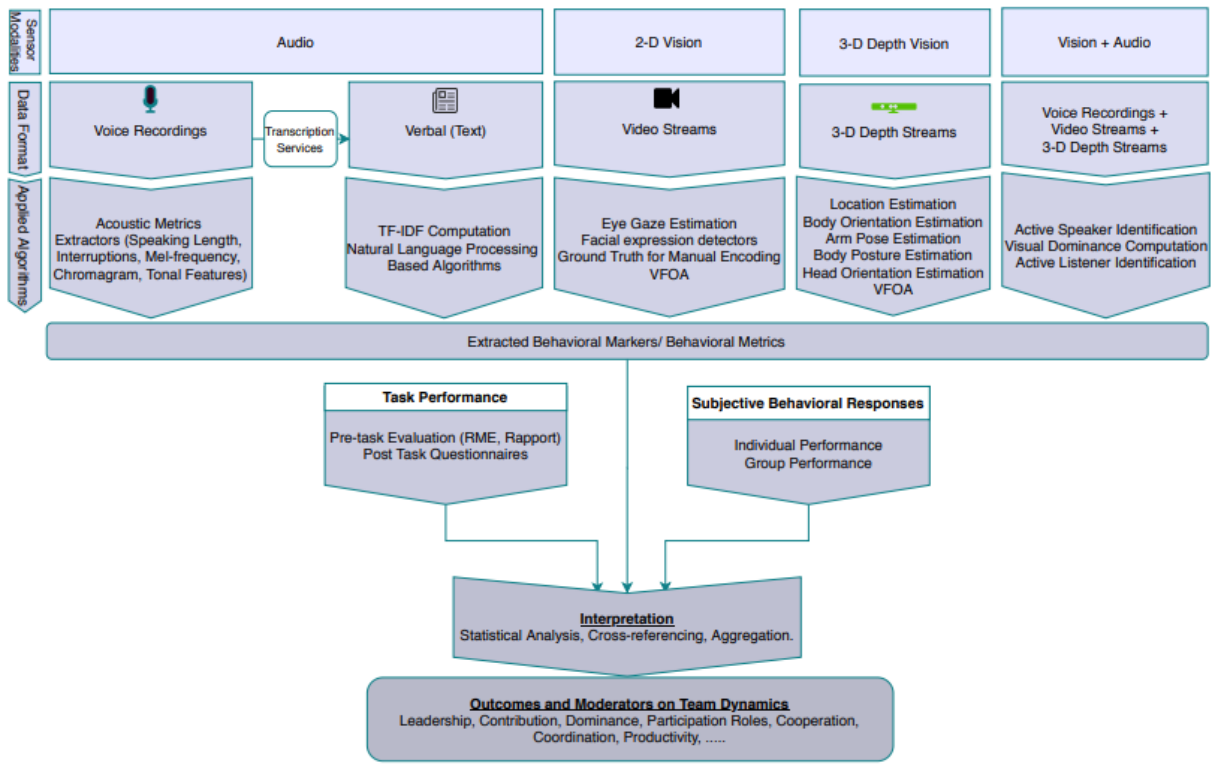

Figure 2: Integration, Illustration of a multimodal analysis framework. Based on different sensor modalities, data is collected in varying formats (audio, text, video, and depth). Signal processing and machine learning algorithms are applied to individual and combined data streams to extract low-level behavioral features that are interpreted via further analysis and user inputs to draw conclusions about group-level outcomes.

**Synchronization steps**

By nature, the output data from the sensors employed for multimodal research is highly heterogeneous, as illustrated in Figure 3. Each of the modalities' output data streams is stored in a different format and requires a different set of processing tools. In addition, the data often needs to be interpreted on different time scales (e.g., 300ms for an audio utterance compared to typical 30 frames per second for video data) which can make the analysis even more challenging. With this in mind, careful synchronization should be performed at all phases of the research study. Synchronization is essential to collective interpretation based on disparate data streams with different frequencies, information content, and formats.

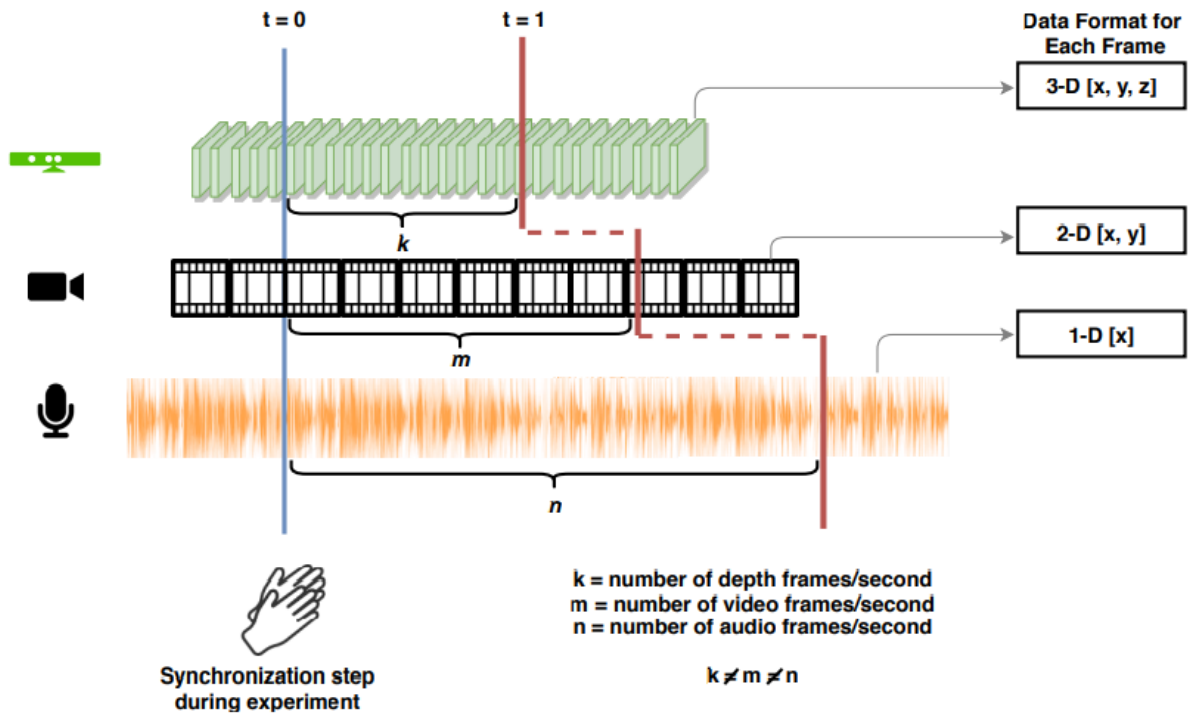

Figure 3 Data Synchronization: Illustration of data streams emanating from different sensor modalities along with a synchronization step. The figure shows that different sensors operate at different frame rates and thus need synchronization so that understanding of time intervals is universal.

A synchronization step is key to establishing a universal time frame as a starting point for all sensors. Any synchronization step needs to be captured by all sensors. As shown in Figure Synchronization, in the case of video, audio, and 3-D sensors, a dramatic clap can act as a universal starting point because it can be picked up by depth and audio-visual sensors. Finally, because the output streams from different sensors operate at different frequencies it is important to synchronize the data frames at regular time intervals and to monitor the set of sensors to ensure recording rates are consistent.

**Feature Extraction**

As shown in Figure 2, each data format (video, 3-D depth, audio, and text) requires processing steps to obtain the behavioral features discussed in earlier sections and outlined in

Figure 1. These steps involve formal algorithms that follow an interdisciplinary approach spanning topics related to organizational research, computer vision, signal processing, machine learning, and natural language processing (Beyan et al., 2016; Beyan et al., 2017a; Beyan et al., 2017b; Zhang & Radke, 2020).

**Audio and Verbal Analysis**

After verbal data is transcribed, natural language processing (NLP) algorithms can be applied to extract useful information (Pilny et al., 2019). Language representation tools like BERT can be applied to extract interpretable relationships in communication patterns from unstructured text (Devlin et al., 2018; Huang et al., 2006). These algorithms are not without problems. Some information that is potentially critical (e.g., talking over each other, backchannels) are usually poorly captured by automatic transcription algorithms and may require considerable manual effort to annotate. In addition, NLP algorithms are frequently trained on, and applied to, text from written corpora (e.g., new articles, tweets) rather than transcriptions of spoken multiparty conversations.

There are fewer reliable off-the-shelf tools to analyze and extract paraverbal interactions than verbal interactions. Some examples of paraverbal extraction software include research that studies signal processing and neural networks applied to audio features which allows researchers to detect emotion from audio alone (Koo et al., 2020; Morgan et al., 2021; Ooi et al., 2014; Zhang et al., 2019). There also exist open-source tools that offer automatic extraction of prosodic features (Huang et al., 2006; Morales et al., 2017).

**Video Analysis**

Video cameras can easily capture non-verbal behaviors such as facial expression, eye gaze, visual focus of attention, micro-level behavioral features and body orientation (Capozzi et al., 2019; Tran et al., 2019; Zhang et al., 2020). Because of its usefulness, analysis of video streams is one of the most pervasive methods of behavior extraction in group research (LeBaron et al., 2018; Waller & Kaplan, 2018). Approaches based on computer vision and machine learning have made classification of human behaviors comparatively easier and accessible to researchers (Waller & Kaplan, 2018; Wu et al., 2017). There are tools that can estimate information such as personality traits (Junior et al., 2019; Okada et al., 2019), leadership (Bhattacharya et al., 2018) and group performance (Murray & Oertel, 2018). Many of these tools are easily accessible through open-source software frameworks like OpenFace (Baltrusaitis et al., 2018).

**3-D Depth Stream Analysis**

Depth information has highly varying levels of complexity. For example, sitting versus standing information can be very different to interpret for algorithms depending on whether the data is depth data vs. image data. A top-down view with depth information can solve this problem through a simple thresholding approach, whereas image-based approaches require a slightly broader multi-step process. Therefore, researchers should carefully consider this complexity when choosing depth sensors and future analysis goals.

Other areas of depth analysis are also current areas of research, and there now exist tools to capture depth information. Researchers have successfully employed a pair of Kinect sensors to record small group interactions and were able to extract various behavioral features such as head pose and orientation, body pose, and arm pose (Bhattacharya et al., 2018; Bhattacharya et al., 2019). In the future there will be doubtless more advances in the realm of depth analysis. For example, gesture and human activity recognition using depth sensors in group interactions is a well-studied problem and holds huge promise for the field of group research (Jalal et al., 2016; Saha et al., 2018).

**Aggregation with Static Data**

Static data is still essential for multimodal group research. Group level and individual behavior are guided by subjectivity and biases that are non-deterministic and therefore hard to incorporate into machine-level understanding. This makes reliance on machines alone for the study of groups limiting or detrimental. To conduct a sound multimodal study, experimenters should cross reference all of their data with the manual behavioral responses of participants through pre/post-test questionnaires.

Post-task questionnaires are designed to obtain aggregate measures of various behavior processes that occurred during group interactions (Bhattacharya et al., 2018; Luciano et al., 2018). They record group and individual experiences, as well as perceptions of each other or the task itself, and are integral to understanding group dynamics. For instance, in the case of studying leadership behavior in small groups, leaders are selected not by appointment but by group perception. From there, experimenters can study the key behavioral features leaders displayed during group interactions to determine leadership qualities.

**Interpretation via Aggregation**

Overall, the goal of multimodal group research is to understand the underlying patterns behind individual and group actions. By aggregating each of the disparate behaviors and we can correlate them to participants' perception of global behavior. The biggest drawback to this method is that the temporality and dynamic ebb and flow are under-represented. Despite this, aggregation is one of the most intuitive ways in which behaviors are interpreted and is instrumental for group dynamics research.

**Incorporating Temporality and Dynamism**

The primary limitation of interpretation through aggregation is the lack of temporal information and contextualization. This is a big limitation; perceptions of group level behaviors such as leadership are very much entangled with the decisions made by participants at different temporal instances (Ericksen & Dyer, 2004; Gerpott et al., 2019; Maruping et al., 2015). In order to truly capture the ebb and flow of group interactions that influence behaviors (McClean, et al., 2019), methods need to constantly consider temporality.

When using a multimodal framework that incorporates temporality, we need to make some tweaks in the analysis approach, as suggested by Kozlowski et al. (2013). The starting framework for this can involve, as shown in Figure 2 and Figure 3, a data collection approach with robust synchronization measures allowing for all modalities of data to be interpretable at uniform time events. This aids in creating intermittent summaries at desired time frames allowing a greater understanding of group processes over time.

**Ethics in Multimodal Research**

Artificial intelligence and machine learning methods have the possibility of both speeding up the experimental and allowing for the measurement of variables that were previously impossible. As useful as these technologies can be, there are also some major drawbacks. There are well known problems of bias in artificial intelligence systems which can come from many sources. For example, the historical data that machine learning systems learn from can be biased (Mehrabi et al., 2021), leading to biased methodology. Part of this difficulty stems from the fact that researchers are often relying on the tools that others have built and therefore must do more work to vet their methods. Because these technologies can improve group communication research, extra care should be taken to ensure that research studies are conducted ethically. In this section, we discuss machine learning algorithm considerations that should be taken in regard to multimodal group research.

Bias here means that the computer algorithm implicitly favors, or disfavors one or more groups as compared to others which can result in faulty conclusions based on biased data. To conduct multimodal research well, we need to ask whether algorithms can do a fair job of extracting key features from our multimodal sources across all the participants we have. For example, face detection algorithms from have over predicted Asians as blinking (Rose, 2010). If a study were to use these algorithms in a study with eye gaze, they might miss critical information about where participants are looking. To examine all the software that currently exists is beyond the scope of this paper and would quickly become obsolete. Instead, we offer our recommendation, which is for researchers to carefully examine the machine learning and AI tools they plan to use.

Our recommendation is that researchers who wish to use machine learning or AI tools when evaluating data should examine whether data used to train models is diverse in ways that are applicable to the study participants. This can mean evaluating the dataset themselves or reaching out to the tools' creators to ask about the training datasets and how they were constructed. Researchers may also request algorithmic auditing.[1] This recommendation is specifically for research regarding in person experimentation, issues with AI in historical data are beyond the scope of this work. This does not mean that researchers need to examine every protected category when using these tools, but rather that all participants will be treated fairly by the algorithms in use. Beyond being an ethical responsibility, this is also good scientific practice.

The extra work required may push some researchers to simply ignore machine learning or AI methods and instead use human coders. We would caution against this. It is not our recommendation that researchers interested in multimodal research simply do not use machine learning or AI methods and instead use human coders. Researchers and assistants are just as prone to the implicit biases that are often present in AI, but because we do not test human coders for their biases this is overlooked. For example, human coders are not always consistent with even themselves (Belur et al., 2021).

Finally, we remind researchers who are considering machine learning algorithms to consider machine learning methods as a tool rather than an objective truth. There is no objective truth and therefore there can be no objective truth for machine learning methods as well.

---

[1] https://www.ajl.org/take-action#REQUEST

Ultimately, this means that we should stop thinking of machine learning as a creator of truth but as a useful tool that can help enhance and speed up the analysis of human interaction.

**Concluding Remarks**

Multimodal research is underrepresented in communication and management research. Our study only found 61 studies from our communication and management journal sample that utilized multiple non-static modalities. Beyond that none of the studies in our sample examined how the different modalities interacted with each other. But from the articles cited here we can see that multimodal group research is being conducted, just not in the communication and management field. Though these are definite drawbacks, the positives of multimodal research far outweigh their negatives. By using a multimodal approach, we can now examine how verbal, para-verbal, visual, depth, and neurophysiological information can influence individual and group outcomes. In addition, we can see how each signal interacts with the others in real time. This gives group researchers a depth of information previously impossible and allows us to come to more complete understandings of group behavior. In communication and management, this could lead to much better understandings of processes like leadership, dominance, bullying, and other multi-feature team interactions. The addition of multimodal research to the communication and management fields would greatly improve their understanding of group processes and human interaction. The requisite tools and prerequisite knowledge are in place for multimodal research. We hope this paper can act as a preliminary guide for communication and management scholars navigating multimodality for the first time.

Appendix 1: Journals that Published Empirical Research in our Sample

Journal of Applied Psychology

Academy of Management Journal

Organizational Behavior and Human Decision Processes

Organization Science

Management Science

Journal of Management

Communication Research

International Journal of Communication

Journal of Communication

Journal of Management Studies

Organization Studies

Administrative Science Quarterly

Journal of Applied Communication Research

Human Communication Research

European Journal of Communication

Organizational Research Methods

Harvard Business Review

Communications-European Journal of Communication Research

Narrative Inquiry

Communication Theory

Visual Communication

Academy of Management Review

1. Coders will need to categorize the papers on a few criteria. First, they will need to categorize the type of paper each article is. The categories are: study, meta-analysis, review paper, and commentary papers.

| Empirical Research | Papers that include empirical research should have an empirical study. They will have a hypothesis and then use observed evidence to support or reject their claims. This includes studies that use surveys, qualitative data etc. |
|---|---|
| Meta-analysis | A meta-analysis is a study that combines the results of many previous studies for increased statistical power. It takes the data of all these studies and combines them to examine whether the sum can tell us more than the parts. |
| Review Paper | A review paper conducts no research and presents a summary of all prior work around a particular topic, like a meta-analysis it looks at many prior works, unlike the meta-analysis it does not rely on the data. |
| Commentary | A commentary paper is an opinion piece by the authors. For example, it could be a logical argument. The authors argue for an idea without empirical evidence using logic and past research. For example, the author might want their field to move in a certain direction and argue for their case. It is not a review paper because review papers examine the state of a topic/field, while commentary papers take a subjective approach of what the author wants.

If the authors propose an equation or model, it is a commentary paper. |

2. After studies have been classified into the aforementioned categories coders will then take papers that were "Empirical Research" and categorize them as multimodal or not.
   a. "Yes" if they are multimodal, "No" if they are not multimodal
   b. Multimodal research for us means that this empirical study used a modality in addition to self-reports to capture information, such as audio, visual or biosensing information. For example, a study is multimodal if it uses a self-report and video cameras to capture data.
3. Studies should be categorized here if they were empirical studies. The categories are: Static Data, Audio, Visual, Biosensing, and every combination of the four.

| Static Data | If participants had to fill out any type of questionnaire or their information was recorded at a specific time point |
|---|---|
| Audio | If participants words or sounds are being recorded in any way E.g. with microphones |

| | Example study (includes 2D vision): Tracking the Leader: Gaze Behavior in Group Interactions |
|---|---|
| 2D Vision | 2D sensors take in the visual information in front of them. E.g. a video camera, still images<br><br>Example study (includes audio as well): Tracking the Leader: Gaze Behavior in Group Interactions |
| 3D Vision | 3D sensors can tell the location of a person's body or gestures.  This includes when people wear sensors that track their location<br>e.g. a depth sensor<br><br>Example study (includes audio as well): A Multimodal-Sensor-Enabled Room for Unobtrusive GroupMeeting Analysis |
| Biosensing | If participants biological signals are being recorded. This includes heart rate or brain monitoring which could be recorded via smart watch or EEG respectively. |

Because there is ambiguity in some papers, these categories should be selected if they are explicitly mentioned in the paper. For example, if a paper mentions that they transcribed a conversation that would not count as audio unless they mention recording.

4.  Only mark if researchers use machine coding and leave blank for other options.

| Machine Coding | This includes whether coding was done with AI or machine learning methods.<br><br>For example, this paper used machine coding to track eye gaze: Tracking the Leader: Gaze Behavior in Group Interactions |
|---|---|